\documentclass[10pt, conference, compsocconf]{IEEEtran}

\usepackage{graphicx}	
\usepackage{color}

\usepackage[dvipsnames]{xcolor}
\usepackage{changes}
	\definechangesauthor[color=red]{LMY}
	\setremarkmarkup{(#2)}

\begin{document}
%
\title{Efficiently Disassemble-and-Pack for Mechanism}


\author{\IEEEauthorblockN{Mingyuan Li}
\IEEEauthorblockA{
Zhengzhou University, ZZU\\
Zhengzhou, China\\
iemyli@gs.zzu.edu.cn}
\and
\IEEEauthorblockN{Xiaoheng Jiang}
\IEEEauthorblockA{
Zhengzhou University, ZZU\\
Zhengzhou, China\\
jiangxiaoheng@zzu.edu.cn}
\and
\IEEEauthorblockN{Ningbo Gu}
\IEEEauthorblockA{
Zhengzhou University, ZZU\\
Zhengzhou, China\\
guningbo@gmail.com}
\and
\IEEEauthorblockN{Weiwei Xu}
\IEEEauthorblockA{
Zhejiang University, ZJU\\
Zhejiang, China\\
weiwei.xu.g@gmail.com}
\and
\IEEEauthorblockN{Junxiao Xue}
\IEEEauthorblockA{
Zhengzhou University, ZZU\\
Zhengzhou, China\\
xuejx@zzu.edu.cn}
\and
\IEEEauthorblockN{Bing Zhou}
\IEEEauthorblockA{
Zhengzhou University, ZZU\\
Zhengzhou, China\\
iebzhou@zzu.edu.cn}
\and
\IEEEauthorblockN{Mingliang Xu}
\IEEEauthorblockA{
Zhengzhou University, ZZU\\
Zhengzhou, China\\
iexumingliang@zzu.edu.cn}
}

\maketitle

\begin{abstract}
In this paper, we present a disassemble-and-pack approach for a mechanism to seek a box which contains total mechanical parts with high space utilization. Its key feature is that mechanism contains not only geometric shapes but also internal motion structures which can be calculated to adjust geometric shapes of the mechanical parts. Our system consists of two steps: disassemble mechanical object into a group set and pack them within a box efficiently. The first step is to create a hierarchy of possible group set of parts which is generated by disconnecting the selected joints and adjust motion structures of parts in groups. The aim of this step is seeking total minimum volume of each group. The second step is to exploit the hierarchy based on breadth-first-search to obtain a group set. Every group in the set is inserted into specified box from maximum volume to minimum based on our packing strategy. Until an approximated result with satisfied efficiency is accepted, our approach finish exploiting the hierarchy.

\end{abstract}


%
\IEEEpeerreviewmaketitle

\section{Introduction}
In this paper, we present a approach to solve the two previously studied problems: mechanism disassembly and bin packing problem \cite{chen2015dapper,attene2015shapes}. In the first step, mechanical object is disassembling to a group set of parts with total minimum volume. It means that joint with \emph{maximum cost} must be disconnect and adjust every parts in each group. In the second step, our approach compute a roto-translation for each of the groups when insert it into specified box. The major contribution of this paper is how to disassembling mechanical object into a group set that can be packed efficiently. Our original contributions can be summarized as follows:
\begin{itemize}
  \item A novel disassembly-and-pack integrated approach to disconnect minimum number of joints and pack efficiency;
  \item A new hierarchical disassembly algorithm that produces a groups set with minimum total volume.
  \item A minimum OBB of mechanical group algorithm that calculate the motion parameters of the parts.
\end{itemize}

\begin{figure}
\centering
\includegraphics[width = 8cm]{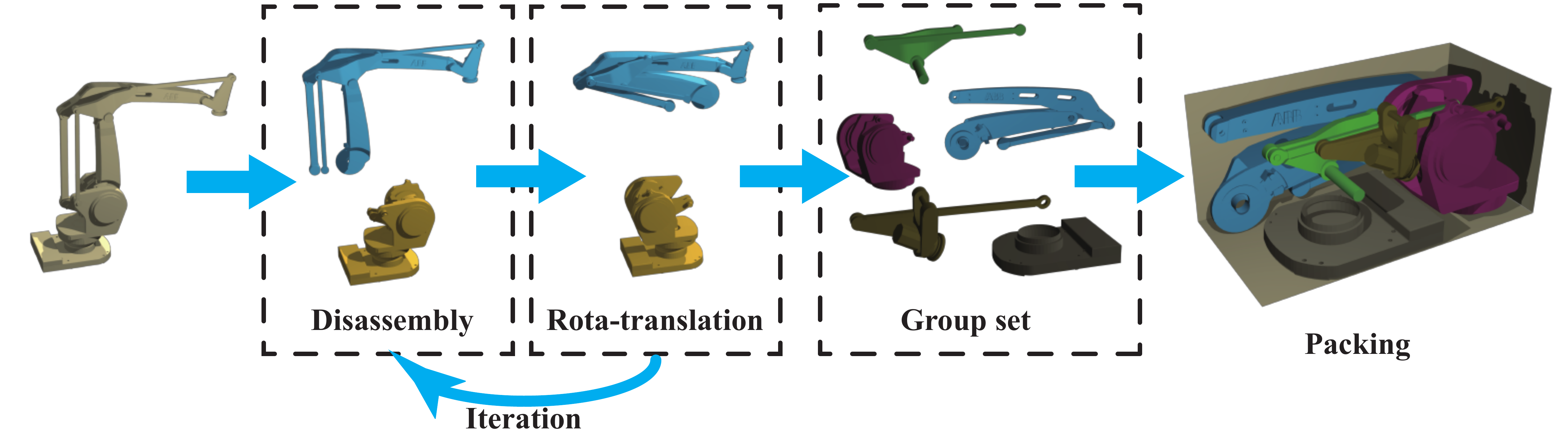}
\caption{\textbf{Overview.} The initial mechanical Robot-arm is disassembled to two groups. Rotate and translate each parts of groups to minimize the volume of every group. Repeat the above two steps to obtain the group set with satisfied disassembling efficiency. Insert the groups into the specified box one by one.}
\end{figure}

\section{Related Work}

 Our problem is a three-dimensions bin packing problem for mechanical object. We hope to minimize the volume of a specified box containing a set of appropriately placed objects. In order to avoid the NP-complete problem \cite{garey1973analysis}, hence heuristic-driven algorithms exist to find approximated solutions. Cutting and Packing approach \cite{chernov2010mathematical}, Decomposing and Packing approach \cite{chen2015dapper} and Splitting and Packing approach \cite{attene2015shapes} can be summarized as two main steps: partition a given object and placing a set of parted objects into a specified container. Our work is mostly related to the Split-and-Pack \cite{attene2015shapes}. However, different from splitting object in \cite{attene2015shapes}, our work focuses on the disassembling mechanical object into parts and adjust each part to minimum volume of OBB for efficiently packing.

\section{Disassemble-and-Pack Approach}

\textbf{Overall Approach:} As the name suggests, the approach consists of two main step: disassembly and packing depicted in Figure 1. The goal of our algorithm is to pack the mechanical object into the specified box with high space utilization. The input to our process includes the mechanical object with its' joint set and a target packing specified box. First of all, our algorithm disassemble mechanical object to group set based on the joint set. Then calculates the parts and rotates and translates parts of every group so that they fit an axis-aligned box of minimum volume. Then in the group number being a constant number of the premise, the algorithm of disassembling expects to get the group set with the minimum sum of group OBB volume. Finally the algorithm efficiently pack each group with rotated and translated into the box in sequence.

\textbf{Hierarchical Disassembling:} Before packing, our approach disassemble the mechanical object to several groups of parts and minimizing the sum of groups MBB volume. We treat a mechanical object as a collection of rigid bodies inter-connected to transmit rigid motions, and a joint connects two parts to form a kinematic pair. Different joint types impose distinct motion constraints between two parts. The relative motion of two parts connected via a joint should be a slippable motion that does not lead to the penetration of parts and it can be determined by the type of their intersection. We summarizes the four main types of joints used in our approach: fixed, revolute joints, gear-2-gear contact joints, and point-on-line joints. The motion of a mechanism is initialized at the driving part and transferred to other parts through its kinematic chain. In a kinematic chain, the motion of a part is restricted to the connected part based on joint type. And it is the summation of transferred motion and \emph{own motion}. For a mechanical part, transferred motion is constant and \emph{own motion} is freedom without the restriction of connected parts.

The hierarchical approach to disassembly mechanical object is popular. In our article, we employ a novel binary hierarchical of disassembling object approach to provide much tighter packing. In each disassembly step, a selected joint is set disconnect and the mechanical object is disassembling to more groups. Then every part in each group has more degrees of freedom to translate and rotate. Since the object volume is constant, minimizing the group box volume means that minimizing the holes volume of the group box. Based on joint type, translate and rotate the parts in each group to minimizing the holes volume. After each disassembly step, we obtain a group set with minimum total volume of group set which will be packing into the box.

\begin{figure}[!t]
\centering
\includegraphics[width = 8cm]{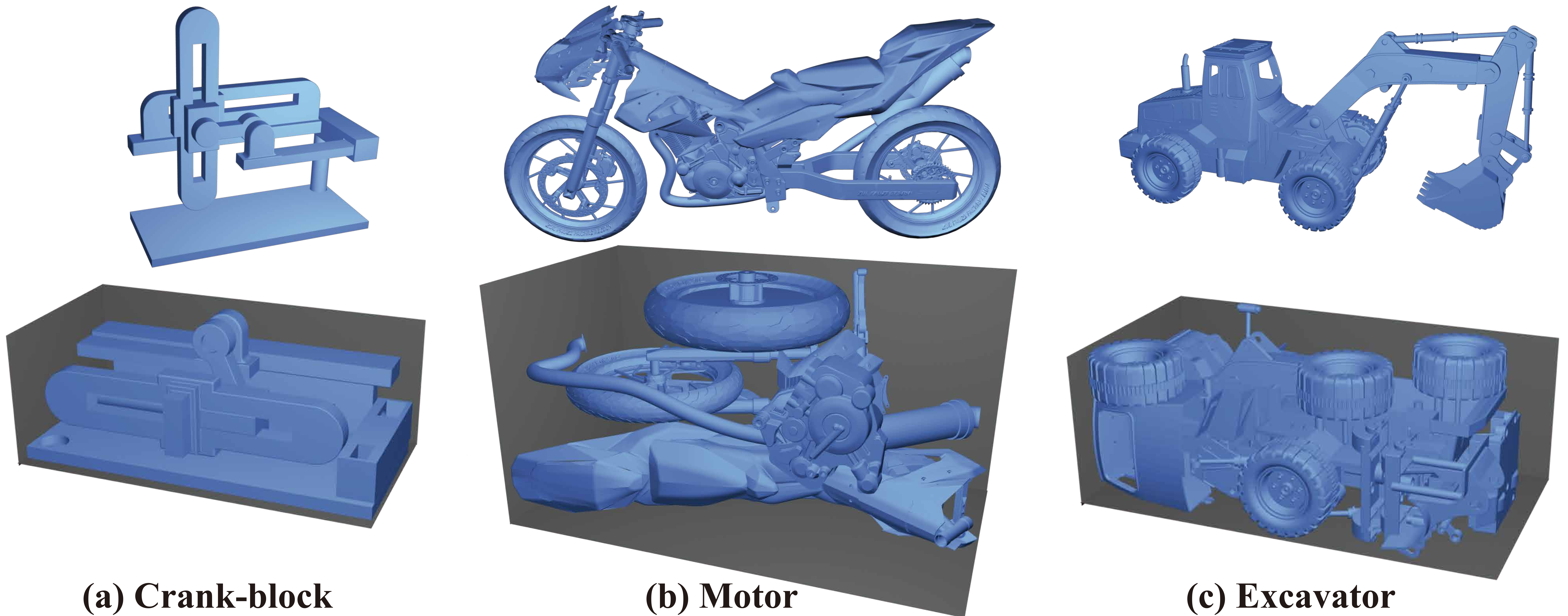}
\caption{Some example models used to test our algorithm.}
\end{figure}

\textbf{Mechanical Group Packing:} After previous section a mechanical object is disassembled to a group set, our objective is to determine a roto-translation for each group so that their overall MBB is minimized. To solve this problem, our approach initialize an axis-aligned box which can contains the all object in group set. Prior to computing such a box, however, the object is rotated according to its MBB, so that the initial box is both axis-aligned and of minimum volume. That means the bottom of the box is constant and the vertical direction of box is variant which decide the volume size. Then groups of parts are inserted into the box one by one while measuring their total height. Minimizing the total height is equal to minimize the total volume. Before insert group into the box, our algorithm sort the groups based on their maximum extension which, for each group, corresponds to the maximum length of its $vol(MBB(g))$. Inserting a group set $G$ into the box is to determine a roto-translation of $g$ like \cite{attene2015shapes}. Our algorithm use quaternion $Q$ and a vector $V$ to defined the rotation and translation of the group, respectively. Simplify the problem of optimal $Q$ and $V$, compute $V$ for a given fixed value of $Q$. The algorithm proceeds as follows: (1) If there are holes that can contain $P$, we select the one whose volume is closer to $P$'s volume and set $V$ accordingly. (2) If no such hole exists, we set $V$ so that $P$'s underlying free volume is minimized while not increasing $h$. (3) If all the positions increase $h$, we set $V$ so that $h$ increases as few as possible.

\section{Results and Discussion}
In this paper, we present a packing algorithm to disassembly a mechanism to a group set and insert them into a box. Showed in Figure 1 and Figure 2, our experiments selected four representative mechanical object: Robot-arm, Crank-block, Motor and Excavator. Experimental results show that our approach can really efficiently pack a range of mechanisms from simple model to complex objects.

\section*{Acknowledgment}
We would like to thank the anonymous reviewers for their constructive comments; Mingliang Xu is partially supported by national science foundation of China (NSFC) (No. 61672469, and No. 61472370). Enjie Ding is partially supported by the national key technology research and development program of China under Grant No. 2017YFC0804401.



%

\bibliographystyle{IEEEtran}
\bibliography{IEEE_Efficient_Packing}

\end{document}